\documentclass[fleqn,usenatbib]{mnras}

\usepackage{newtxtext,newtxmath}

\usepackage[T1]{fontenc}

\DeclareRobustCommand{\VAN}[3]{#2}
\let\VANthebibliography\thebibliography
\def\thebibliography{\DeclareRobustCommand{\VAN}[3]{##3}\VANthebibliography}

\usepackage{graphicx}	
\usepackage{amsmath}	
\usepackage{xcolor}     

\usepackage{comment}
\newcommand{\VEC}[1]{\vec {#1} }

\title[Relic of deformation in a neutron star]{Relic of quadrupole deformation produced in a hot neutron star era}
\author[Y. Kojima et al.]{Yasufumi Kojima,$^{1}$\thanks{E-mail: ykojima-phys@hiroshima-u.ac.jp}
 Akira Dohi, $^{2,3}$
 Shota Kisaka,$^{1}$
 and Kotaro Fujisawa$^{4}$
\\
$^{1}$Department of Physics, Graduate School of Advanced Science and Engineering,Hiroshima University, Higashi-Hiroshima 739-8526, Japan\\
$^{2}$Astrophysical Big Bang Laboratory (ABBL), Cluster for Pioneering Research, RIKEN, Wako, Saitama 351-0198, Japan \\
$^{3}$Interdisciplinary Theoretical and Mathematical Sciences Program (iTHEMS), RIKEN, Wako, Saitama 351-0198, Japan \\
$^{4}$Department of Liberal Arts, Tokyo University of Technology, Ota, Tokyo 144-8535, Japan\\
}

\date{Accepted XXX. Received YYY; in original form ZZZ}

\pubyear{\the\year{}}

\begin{document}
\label{firstpage}
\pagerange{\pageref{firstpage}--\pageref{lastpage}}

\maketitle

\begin{abstract}
A newly born neutron star is expected to exhibit significant deviations from spherical symmetry, which decay with time.
Determining how much deformation remains at present is crucial for gravitational-wave astronomy. This study is the first investigation into the evolution of quadrupole deformation during the solid crust formation phase to obtain a plausible value at present.
The equilibrium structure before solidification is modeled using a fluid description, and the deformation is introduced through an assumed driving force.
As the star cools, this force weakens, leading to a gradual decay of the deformation. Eventually, the deformation vanishes in the fluid region but partially remains in the crust, sustained by elastic forces, after solidification. By comparing the equilibrium models before and after solidification, we estimate the residual ellipticity and
demonstrate that the spatial profile of the elastic shear is imprinted in the crust.
The relic ellipticity is only a few percent of the original value, with its absolute magnitude depending on the deformation mechanism during the hot era, which cannot be specified owing to the lack of elaborate models.
This work provides a first step toward linking early neutron star deformation with future gravitational-wave observations.

\end{abstract}


  \section{Introduction}
A time-varying non-axisymmetric mass distribution in spinning neutron stars continuously emits gravitational waves at a fixed frequency~\citep[][]{1979PhRvD..20..351Z}.
These types of gravitational waves are called continuous gravitational waves (CGWs). Currently, CGWs are one of the most promising targets in gravitational-wave astronomy, following the successful detection of transient bursts from compact star mergers~\citep[][]{
2016PhRvL.116f1102A,2016PhRvL.116x1103A,2017PhRvL.119p1101A,2017ApJ...848L..12A,2021ApJ...915L...5A}.
However, no evidence of such CGW signals exists so far~\citep[][for recent result]{2025ApJ...983...99A}.
The CGW amplitude depends on the magnitude of the non-axisymmetric deformation. As such, the upper limits on the signal amplitude are set based on the ellipticity $\epsilon$, 
which quantifies the asymmetry in the neutron star mass distribution.
Current results from the LIGO-Virgo-KAGRA Collaboration, a global network of ground-based gravitational-wave detectors~\citep[][]{2025ApJ...983...99A}, show that,
for 29 pulsars, the upper bound is less than the spin-down limit, 
$\epsilon = 2\times 10^{-4} (P/0.1{\rm{s}})^{3/2}(\dot{P}/10^{-15})^{1/2}$,
which is the maximum ellipticity derived using the quadrupole radiation formula based on the
observed spin $P$ and its time derivative $\dot{P}$.
CGW detection provides crucial insights into neutron star astrophysics and the properties of high-density matter in their interiors.
Possible CGW emission mechanisms,  experimental data analysis, and its history have been reviewed 
in detail~\citep[e.g.][]{2023APh...15302880W,2023LRR....26....3R}.
An axisymmetric star about its spin axis has no time-varying quadrupole moment and hence, does not radiate gravitational waves.
The structure of a rotating fluid star is axisymmetric; hence, the non-axisymmetric deformation must be due to some additional mechanisms.
A possible candidate is the intense magnetic field inside the neutron star
\citep[e.g.,][]{1996A&A...312..675B,1999A&A...352..211K,2021EPJA...57..249C} because the magnetic deformation direction is independent of the spin direction.
The ellipticity $\epsilon$, which characterizes the quadrupole deformation, is estimated by
$\epsilon\sim 10^{-9} (\langle B \rangle/10^{13}G)^2$, where
$\langle B \rangle$ is the average of the field strength inside stars.
For a magnetic deformation with a superconducting core, 
$\epsilon$ is given by
$\epsilon\sim 10^{-8} (\langle B \rangle/10^{12}G)(\langle H_{c} \rangle/10^{15}G)$
\citep{2012MNRAS.419..732L,2014MNRAS.437..424L}, where
$ H_{c}$ is the lower critical field strength for type II superconductivity.
\citet{2018ApJ...863L..40W} suggested an ellipticity of $\epsilon \sim 10^{-9}$,
commonly maintained in millisecond pulsars, to account for
the population cutoff on the $P$-${\dot{P}}$ diagram of the pulsars.
The minimum ellipticity may originate from the magnetic deformation of the superconducting core despite their weak surface fields.
Accreting matter along the
magnetic field lines of a weakly magnetized neutron star in a binary system can also form a small ''mountain''
\citep{2001PASA...18..421M,2004MNRAS.351..569P,
2019MNRAS.484.1079S,2020MNRAS.493.3866S,
2022MNRAS.516.5196F,2023MNRAS.526.2058R,2025ApJ...979...10R}.
The solid crust is also another possibility that can sustain a small irregularity, such as the ''mountain'' on a neutron star.
The crust fractures when the deforming force is too strong. 
Thus, the deformation is determined by the ratio of the elastic force to gravity.
The maximum ellipticity of the solid crust is estimated by
shear modulus $\mu$ and the maximum strain $\sigma_{\rm{max}}$;
$\epsilon \sim $ $(\Delta M/M)\times $ 
$(\mu \sigma_{\rm{max}}/\Delta r)/( GM\rho_{\rm{crust}}/R^2) \sim $
$10^{-6} (\sigma_{\rm{max}}/0.1)$,
where
$\Delta M$ is the crustal mass, $\Delta r$ is the thickness, and
$\rho_{\rm{crust}}$ is the mass density~\citep[see][for detailed calculations]{2000MNRAS.319..902U,2006MNRAS.373.1423H,2013PhRvD..88d4004J}.
The maximum value is attained for all possible forces to produce the mountains.
The nature and extent of the force relevant to the maximum is not clear.
The actual ellipticity is to be smaller and to depend on the explicit mechanism, which is uncertain at present.
The temperature anisotropies in the accreted crust
\citep{2000MNRAS.319..902U}, for which the resultant ellipticity is 
$\epsilon = 2.8 \times 10^{-6}$, are one of the convincing mechanisms.
See also the recent detailed calculations of the thermal mountain
\citep{2023MNRAS.522..226H,2025MNRAS.540.2349J}.
\citet{2021MNRAS.500.5570G} recently proposed a novel formalism to satisfy the boundary conditions at the crust--core interface,
which were incorrectly investigated in a previous study.
In their approach, the authors calculated the deformations due to an introduced force
in fluid and fluid--elastic systems.
The deformation sustained by the elastic force was determined via subtraction.
Their results can be represented as $\epsilon =10^{-8} - 10^{-7}$. These values are significantly lower than the maximum value $\epsilon \sim 10^{-6}$
estimated in a previous work.
However, a different choice of the driving force leads to
$\epsilon = 7.4 \times 10^{-6}$~\citep{2022MNRAS.517.5610M}.
Our concern is a plausible value of $\epsilon$,
although the elastic force sustains a maximum deformation of $\epsilon \sim 10^{-6}$.
We consider an evolutionary scenario that leads to a deformation relevant to
future observations.
In the history of a newly born neutron star, the hot stage is very energetic, and various hydrodynamical irregularities are likely to appear.
The crustal region is assumed to deviate from spherical symmetry owing to various perturbing forces in the hot era.
Their fates in a later period are interesting but cannot yet be simulated in proto-neutron star evolution models that assume spherical symmetry. 
Determining whether the deformation at birth vanishes
around the crust formation time is crucial, with significant implications for future investigations. 
The dynamical force balance can be modeled by a fluid description above the solidification temperature $T_c$. 
The critical temperature is approximately given by
$T_{c} \approx 6.4\times 10^{9} (\rho/10^{14} {\rm{g~cm}}^{-3})^{1/3}$K~\citep{1983bhwd.book.....S}.
When the star cools at time $\sim 100$ s after its birth, 
a solid crust forms around the core--crust boundary and expands radially 
according to the ratio of the Coulomb energy of the nuclei to the thermal energy
\citep[][]{2014PASJ...66L...1S,2022MNRAS.511..356P,2025PTEP.2025l3E01N}. 
The hydrodynamical equilibrium state also changes gradually, and an elastic force acts on the new state after the solid crust forms.
The irregularities caused by thermal energy are washed out in the fluid region, whereas a part of them remains in the crust, sustained by the elastic force.
Simulating the time evolution of the deformation that occurred in the past is challenging. Instead, we consider the structural variation over time by comparing the equilibrium models at different evolutionary stages. A similar idea has already been employed to calculate the elastic strain due to the change in the centrifugal force as a model of pulsar glitches \citep[e.g.][]{2000ApJ...543..987F,2020MNRAS.491.1064G}.
This approach is explained in Fig.~1 in \citet{2021MNRAS.500.5570G}.
However, our formalism only concentrates on the crust part. As such, it
differs from that of \citet{2021MNRAS.500.5570G}
on several technical points.
Our consideration is limited to the quadrupole deformation, which is relevant to CGWs.
The remainder of this paper is organized as follows:
Section 2 describes the mathematical formulation and assumptions for the crustal model.
The deformation during the hot stage of the neutron star is modeled by a driving force, which
is introduced by a working hypothesis.
However, the deformation is not maintained owing to the vanishing force. 
We estimate the residual deformation by comparing dynamical equilibrium models at different times.
The explicit model for the deformed model is given in Section 3.
In Section 4, the residual deformation sustained by the elastic force is calculated.
In Section 5, we discuss the residual quadrupole moment, which is important for gravitational-wave physics.
Finally, our conclusions are presented in Section 6.

  \section{Mathematical formulation}
  \subsection{Crust as a background model}
We consider the crust of neutron stars, whose mass densities range from $\rho_{c}$ 
$=1.4\times 10^{14}$ g cm $^{-3}$ to the neutron-drip density $\rho_{1}=4\times 10^{11}$ g cm$^{-3}$.
The inner radius is $r_{c}$ at the core--crust interface, and the outer radius is chosen as $R$.
We ignore an extremely thin outer crust corresponding to the low density;
$R$ is the stellar surface.
The thickness is assumed to be $\Delta r/R=(R-r_{c})/R=0.1$.
In addition, we ignore the magnetic field and stellar rotation (they are important elements for realistic models 
but significantly complicate the analysis at the same time).
The hydrostatic equilibrium under Newtonian gravity is determined as follows:
\begin{equation}
    p^{\prime} = -\frac{GM_{r} \rho }{r^2}
  \approx    -\rho g_s\left(\frac{R}{r}\right)^2,
\label{hydroequil.eqn} 
\end{equation}
where $M_{r}$ is the mass within a radius $r$.
The crustal mass is extremely small compared with the total mass $M$; hence, 
$M_{r}$ can be approximated as $M$.
The gravitational acceleration $g_s$ at the surface is chosen as $g_s = GM/R^2 \approx 1.5 \times 10^{14}$ cm s$^{-2}$
for $M \approx 1.5 M_{\odot} $ and $R\approx 12$ km.
In the realistic equilibrium model, the adiabatic index $ \Gamma = d\ln p_0/ d\ln \rho_0 $ is a function of density. 
However, we approximate it as a constant in space and use $\Gamma = 3/2$ 
adequate for a high-density region, which is important for mass moment~\citep[see realistic result in Figure~32 of ][]{2008LRR....11...10C}.
The polytropic approximation yields the analytic solution of eq.~(\ref{hydroequil.eqn}) as
\begin{align}
    \rho_{0} & =\rho_{c}\left[1+\alpha
   (r^{-1}-r_c ^{-1} )\right]^2,
   \label{polytropic1.eqn} 
   \\
     p_{0} & =p_{c}
     \left[1+\alpha
   (r^{-1}-r_c ^{-1} )\right]^3,
\label{polytropic2.eqn} 
\end{align}
where $\alpha = (1-(\rho_{1}/\rho_{c})^{1/2})r_c R/\Delta r$. The pressure $p_c$ at $r=r_c$ is given by
 $p_c  = \rho_c g_s R^2 /(3\alpha)$  $\approx 9.8 \times 10^{32}$ dyn cm$^{-2}$, and radially decreases 
$p_{0}(R) =p_{c}(\rho_{1}/\rho_{c})^{3/2}\approx 1.5 \times 10^{-4} p_{c}$
$\approx 1.5 \times 10^{29} $ dyn cm$^{-2}$.
The crustal mass $\Delta M$ is determined to be extremely small
$\Delta M = 4\pi \int \rho_{0} r^2 dr $ $\approx 3 \times 10^{-2} M$,
integrating it from $r_c$ to $R$.
We introduce a sound velocity $c_s$ for the spatial profile of eqs.~(\ref{polytropic1.eqn})--(\ref{polytropic2.eqn});
\begin{equation}
    c_s = \left( \frac{p_{0} ^\prime}{\rho_{0} ^\prime}
    \right)^{1/2} .
    \label{def_cs.eqn}
\end{equation}
This ranges from $c_s =3.2 \times 10^{9}$cm s$^{-1}$ at $r=r_c$
to $7.5 \times 10^{8}$cm s$^{-1}$ at $R$.
The shear modulus $\mu$ in the solid crust may be
approximated as a linear function of density,
which is overall-fitted to the results of a detailed calculation reported in a previous study~\citep[see Figure~43 in ][]{2008LRR....11...10C}.
Thus, $\mu$ is given by a radial function, approximated
in terms of the crustal spatial density profile
 \begin{equation}
    {\mu} = \mu_{c}
    \left[1+\alpha(r^{-1}-r_c ^{-1} )\right]^2,
\label{DFshMD.eqn} 
 \end{equation}
where $\mu_{c}=10^{30}~{\rm erg~cm}^{-3}$ at the core--crust interface.

  \subsection{Deformed crust}
We consider a small deviation from the spherical crust modeled by $\rho_0(r)$ and $p_0(r)$ (eqs.~(\ref{polytropic1.eqn})--(\ref{polytropic2.eqn})).
A deformed model is not unique owing to variety resulting from the incorporation of physical properties.
Our formalism with assumptions is discussed below. 
We consider the radial region ($r_c\le r \le R$), which is assumed to be fixed for simplicity, i.e., we neglect the radial perturbation with $l=0$.
Consider an equilibrium state deformed by a force density $\VEC{f}$ in a hot fluid state.
For $T> T_{c}$, the solid crust does not form; however, the corresponding density region 
($r_c\le r \le R$) is considered. 
The magnitude $|\VEC{f}|$ is sufficiently small to use the linearized perturbation method.
The deformed state is given as A, and the variation from the background model
is written with a subscript ''a,'' as in $\delta p_a$.
The deformed equilibrium by ${\VEC{f}}$ is governed by
\begin{align}
& {\VEC \nabla} \delta p _{a} +\delta \rho _{a} {\VEC g} 
+ \rho_{0} {\VEC \nabla} \delta \Phi_{a} 
=\VEC{f},
\label{balanceBF.eqn}
\\
& \nabla ^2 \delta \Phi_{a} = 4 \pi G_{\rm{N}} \delta \rho_{a} ,
\label{PoissonA.eqn}
\end{align}
where $\VEC{g} \equiv {\VEC \nabla} \Phi_{0}$.
The relation between $\delta p_a$ and $\delta \rho_a$ in state A
is not $\delta p_a = c_s ^2\delta \rho_a$, but
\begin{equation}
    \delta p_a = c_s ^2\delta \rho_a + Q.
    \label{relationA.eqn}
\end{equation}
The force balance condition with ${\VEC{f}}$ does not necessarily result in the proportional relationship between 
$\delta p_{a} $ and $\delta \rho_{a}$.
The ${\VEC{f}}$-induced variation in pressure is generally associated with that in the density and other parameters, e.g., 
 the entropy per baryon $s$ for $p=p(\rho,s)$.
The form (\ref{relationA.eqn}) implies that state A is not related to the adiabatic change from the background state.
The state is sufficiently hot, and non-spherical irregularities are excited by a large thermal energy, which is smaller than the gravitational energy. As such, perturbation is applicable. 
The term $Q$ is related to $\VEC{f}$.
We discuss $\VEC{f}$ and $Q$ for an explicit model, and the solution of eqs.~(\ref{balanceBF.eqn})--(\ref{PoissonA.eqn}) in the next section.
The deformation is not fixed in time.
Over a long period, the force gradually decays and eventually vanishes with decreasing thermal energy.
The equilibrium structure changes accordingly, with the new state given as B.
The equilibrium structure after $\VEC{f}=0$ is reached can still be described as a small deviation from that of the spherical crust.
This is denoted by a sum, such as $\delta p_{a} + \delta p_{b}$.
An elastic force acts to ensure that the force is balanced after solidification; hence, the perturbation equation in state B is 
\begin{equation}
{\VEC \nabla} (\delta p_{a} + \delta p_{b} )
+ (\delta \rho_{a} + \delta \rho_{b} ){\VEC g} 
+ \rho_{0} {\VEC \nabla} (\delta \Phi_{a} +\delta \Phi_{b})  
+ {\VEC{h}} =0.
\label{balanceEND.eqn}
\end{equation}
The elastic force ${\VEC{h}}$ is given by
\begin{align}
   & h^{i} = -\nabla_{j} ( 2\mu\sigma^{ij} ),
    \\
& \sigma_{ij} =\frac{1}{2}
\left( \nabla_{i} \xi_{j}+
\nabla_{j} \xi_{i}  \right)
-\frac{1}{3} g_{ij} \nabla_{k} \xi^{k} .
\end{align}
Consider a limiting case, in which the elastic force is exactly zero.
Equation~(\ref{balanceEND.eqn}) is the same as 
eq.~(\ref{balanceBF.eqn}) without the source term${\VEC{f}}$.
The non-zero solution of eq.~(\ref{balanceBF.eqn}) is determined by 
the source term. Therefore, the solution of eq.~(\ref{balanceEND.eqn}) 
for the sourceless case is given by
$\delta p_{a}+\delta p_{b}$
$=\delta \rho_{a}+\delta \rho_{b} $
$=\delta \Phi_{a}+\delta \Phi_{b} =0$.
However, ${\VEC{h}}$ in eq.~(\ref{balanceEND.eqn}) is generally a non-zero term because
the displacement $\VEC{\xi}=0$ in state A changes to $\VEC{\xi} \ne 0$ in state B.
Therefore, we expect 
$\delta p_{b} \ne-\delta p_{a} $,
$\delta \rho_{b} \ne -\delta \rho_{a}$, and
$\delta \Phi_{b} \ne -\delta \Phi_{a}$.
An elastic deformation generally remains in state B.  
By subtracting eq.~(\ref{balanceBF.eqn}) from eq.~(\ref{balanceEND.eqn}), we have 
\begin{equation}
{\VEC \nabla}  \delta p_{b}
+\delta \rho_{b} {\VEC g} 
+ \rho_{0} {\VEC \nabla} \delta \Phi_{b}  
+ {\VEC{h}} 
=-\VEC{f}.
\label{balanceAF.eqn}
\end{equation}
The change in the gravitational potential $\delta \Phi_{b}$ is governed by $\delta \rho_{b}$;
\begin{equation}
\nabla ^2 \delta \Phi_{b} =4 \pi G_{\rm {N}} \delta \rho_{b} .
\label{PossionPT.eqn}
\end{equation}
The Eulerian change in density is derived by the mass conservation
\begin{equation}
\delta \rho_{b} = -
{\VEC{\nabla}}\cdot (\rho_{*} {\VEC{\xi}}) 
\approx -
{\VEC{\nabla}}\cdot (\rho_{0} {\VEC{\xi}}) ,
\label{densitybyxi.eqn}
\end{equation}
where $\rho_{*}$ is the mass density in state A, and is only different from
$\rho_{0}$ by a small quantity of the first order.
By neglecting small quantities in the second order, we approximate $\rho_{*}$ as $\rho_{0}$ in eq.~(\ref{densitybyxi.eqn}).
The adiabatic relation is assumed to be satisfied between the density and pressure deviations in state B from the background model;
\begin{equation}
\delta p_{a} +\delta p_{b} =c_s ^2
  (\delta \rho_{a} +\delta \rho_{b}) .
\end{equation}
By using eq.~(\ref{relationA.eqn}), the change from state A to B is given by
\begin{equation}
\delta p_{b} =c_s ^2 \delta \rho_{b} -Q.
\label{presbwithQ.eqn}
\end{equation}
In summary, the deformed structure in a fluid system is determined by eqs.~(\ref{balanceBF.eqn})--(\ref{PoissonA.eqn}), with force density $\VEC{f}$.
The change in the shrinking phase after solidification is determined by eqs.~(\ref{balanceAF.eqn})--(\ref{PossionPT.eqn})
with $ -\VEC{f}$.
Relic deformation is the sum of the initial deformation and its shrinkage, e.g.,
$\delta \rho_{a} +\delta \rho_{b}$.

  \section{Deformation in the hot state}
\begin{figure}\begin{center}
  \includegraphics[width=0.75\columnwidth]{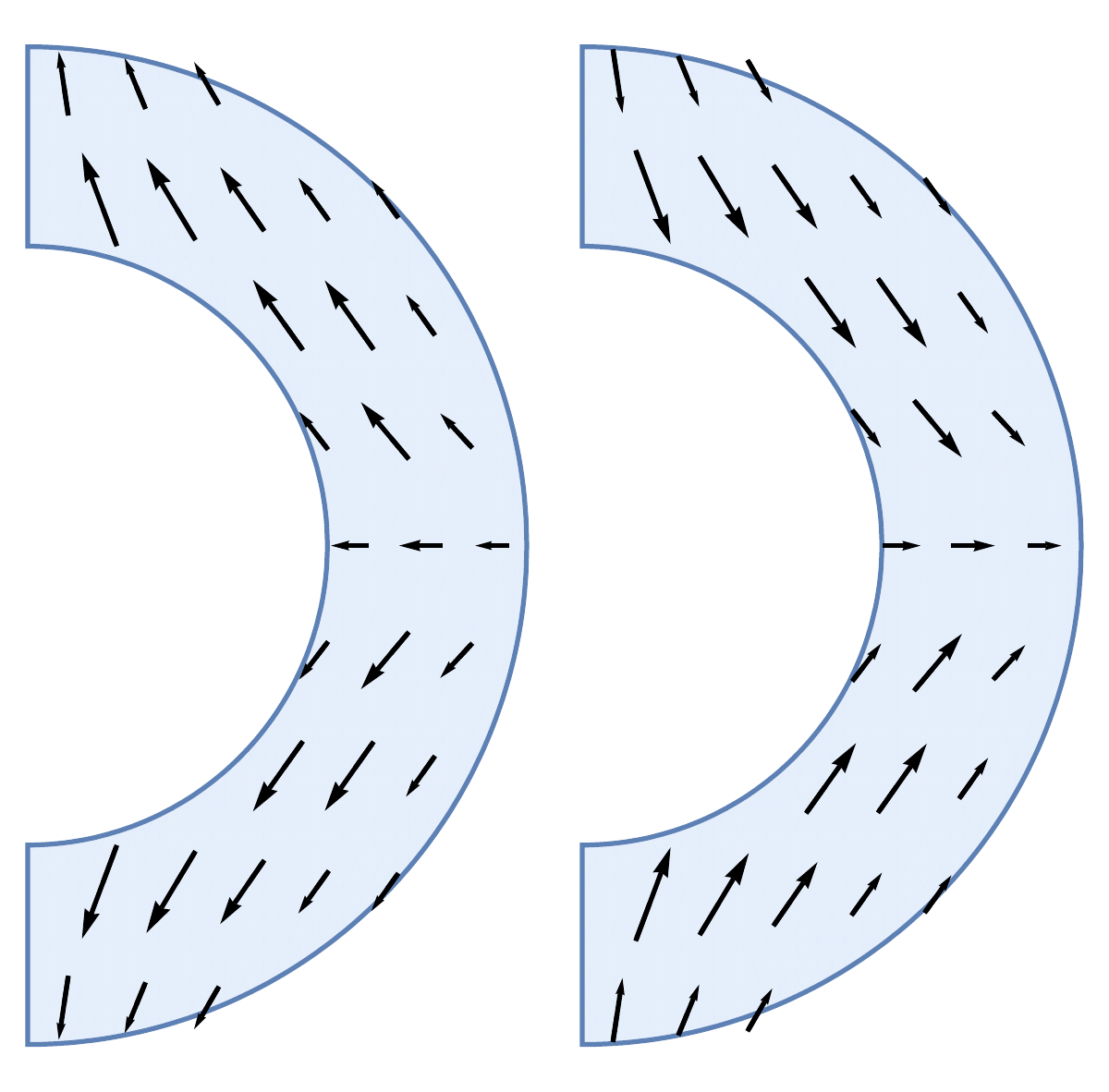}
\caption{ 
 \label{Fig1}
The force density vector ${\VEC{f}}$ is schematically described by arrows in the meridian plane of the crust.
Positive functions ($A_2>0, B_2>0$) in eq.~(\ref{explictradfnH.eqn}) result in prolate deformation, as shown in the left panel, whereas negative functions ($A_2<0, B_2<0$) oblate in the right panel.
}
\end{center}\end{figure}

We consider an axially symmetric model of the deformed crust.
The driving force is uncertain 
owing to the lack of information on the hot state at birth.
Possible terms appearing in the force density $\VEC{f}$ in the hot era are 
$\rho({\VEC{v}_{P}} \times ({\VEC{\nabla}}\times {\VEC{v}_{P}}) -{\VEC{\nabla}}v_{P}^2/2)$, which are attributed to
the circulation flow, or the Lorentz force
$({\VEC{\nabla}}\times {\VEC{B}_{P}})\times {\VEC{B}}_{P}/(4\pi)$ .
They are assumed to be significantly dissipated to a negligible level $\VEC{f} \approx 0$ in the late state. 
When the poloidal flow velocity ${\VEC{v}_{P}}$ or 
the poloidal magnetic field ${\VEC{B}_{P}}$ is dipolar ($l=1$),
$\VEC{f}$ incorporates the quadrupole ($l=2$) part in the angular dependence.
The explicit form is written according to the Legendre polynomial
$P_l(\cos\theta)$ with $l=2$;
\begin{equation}
   \VEC{f}= \rho_{0}( A_2 P_2{\VEC{e}}_r+
        B_2 P_{2,\theta} {\VEC{e}}_\theta ) ,
\label{explictradfnH.eqn}
\end{equation}
where $A_2$ and $B_2$ are radial functions. 
In our approach, we mathematically model them in the following way.
Both functions are arbitrary; however, they are chosen in the same form for simplicity.
The explicit form is as follows:
\begin{align}
&A_2 = B_2 = N_n(r-r_c)^2(r-R)^2~~~~~~~~~~~~~~~~~~~~~~~~~~~~~~~~~~~~(\text{Model I}), 
\\
&A_2 = B_2 = N_n(r-r_c)^2(r-R)^2\exp[-(w(r-r_{92}))^2] ~~(\text{Model II}), 
\\
&A_2 = B_2 = N_n(r-r_c)^2(r-R)^2\exp[-(w(r-r_{98}))^2] ~~(\text{Model III}), 
\end{align}
where $w=(0.1\Delta r)^{-1}$, $r_{92}=r_c +0.2\Delta r$,
$r_{98}=r_c +0.8\Delta r$, and 
$N_{n}$ is the normalization constant.
These radial functions are chosen such that
the perturbation is localized inside the crust, i.e., the radial functions both approach zero at the boundaries.
In models II and III, an exponential function is used to ensure further localization near the core or surface. 
The vector $\VEC{f}$ is illustrated by arrows in Fig.~\ref{Fig1}.
The figure shows that the quadrupole deformation is prolate
for $A_2>0, B_2>0$, whereas it is oblate for $A_2<0, B_2<0$.
%

\begin{figure}\begin{center}
\includegraphics[width=1.0\columnwidth]{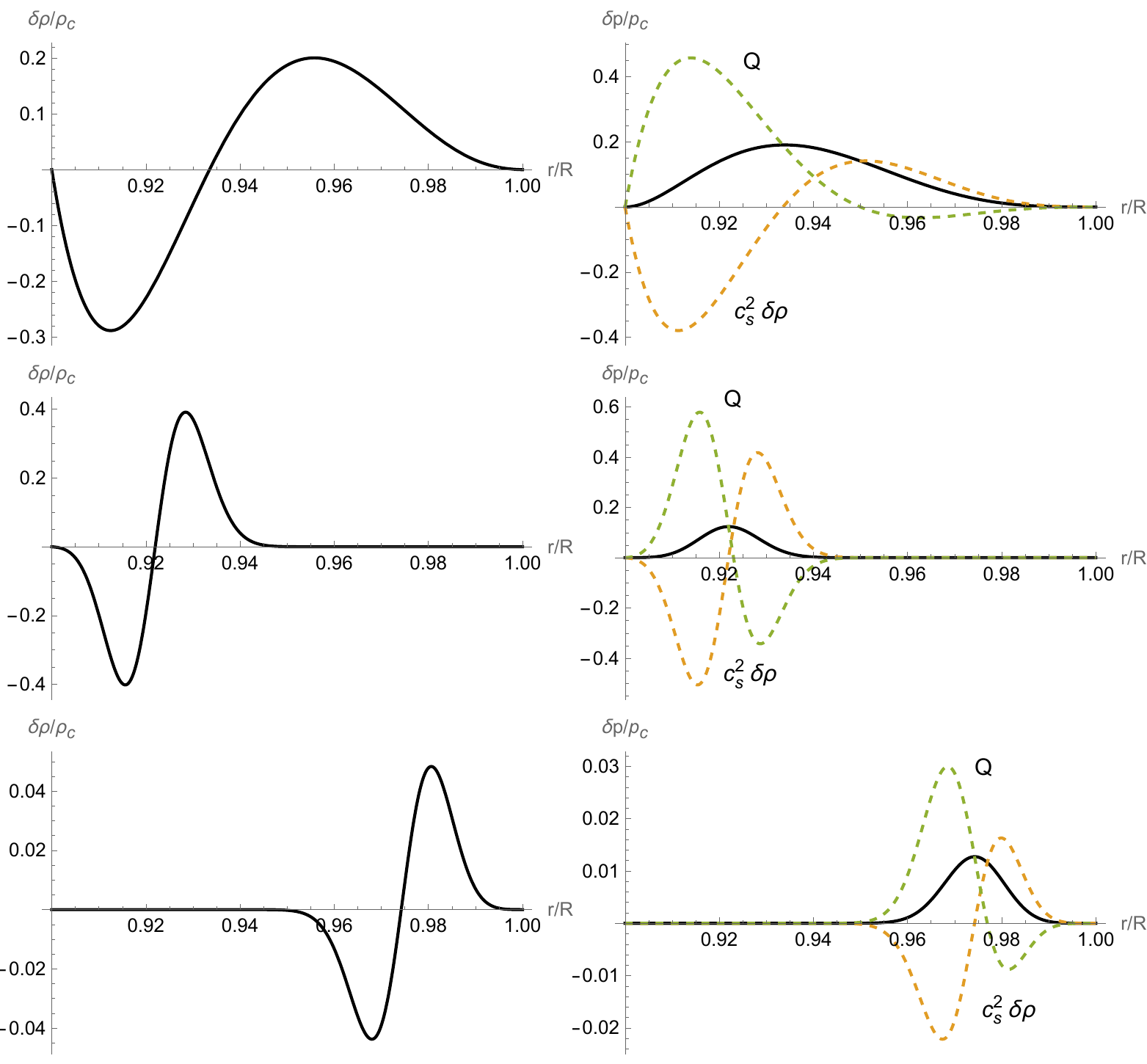}
\caption{ 
\label{Fig2}
 Density perturbation $\zeta^{-1}\delta \rho_{a2}$ (left panel) 
and pressure perturbation $\zeta^{-1}\delta p_{a2}$ (right panel)
are shown as a function of the radius for Models I--III from top to bottom.
In the right panels, 
$\zeta^{-1}Q_{2}$ and $\zeta^{-1}c_s ^2 \delta \rho_{a2}$ are shown using dashed curves.
}
\end{center}\end{figure}

We solve eqs.~(\ref{balanceBF.eqn})--(\ref{PoissonA.eqn})
by expanding $\delta p_a$, $\delta \rho_a$, and $\delta \Phi_a$ using
 the Legendre polynomial, e.g., 
$\delta p_a$ $=\sum \delta p_{al} (r)P_l(\cos\theta)$.
The functions with each index $l$ are decoupled.
Henceforth, we only consider the $l=2$ part relevant to quadrupole deformation.
The radial and its perpendicular components in eq.~(\ref{balanceBF.eqn}) are reduced to
\begin{align}
   \delta \rho_{a2} &= \frac{1}{g} 
   \left(\rho_0 ^\prime
   \delta \Phi_{a2} +\rho_{0}A_2-(\rho_{0}B_2 r)^\prime 
   \right).
   \label{fld_dns_phi.eqn}
   \\
  \delta p_{a2} &= -\rho_0 ( \delta \Phi_{a2} -B_2 r),
\label{fld_pr_phi.eqn}
\end{align}
By eliminating $\delta \Phi_{a2}$ in eqs.~(\ref{fld_dns_phi.eqn})--(\ref{fld_pr_phi.eqn}), the relation between $\delta p_{a2} $ and $\delta \rho_{a2}$ is written as
\begin{equation}
  \delta p_{a2} = c_s ^2 \delta  \rho_{a2} + Q_{2}, 
    \label{p2withq2.eqn}
\end{equation}
where $c_s$ (eq.~(\ref{def_cs.eqn})) and 
$p_0 ^\prime =-\rho_0 g$ are used, and $Q_{2}$ is given by
\begin{equation}
  Q_{2} = \frac{\rho_0 ^{2}}{\rho_0 ^\prime}
  \left[ A_2-({B_2 r})^\prime
  \right].
  \label{def_q2.eqn}
\end{equation}
The relation (\ref{p2withq2.eqn}) is derived from the force balance equation.
The thermodynamical relationship between 
$\delta p_{a}$ and $\delta \rho_{a}$ is no longer specified redundantly,
and additional equations would constrain the force density $\VEC{f}$.
The term $Q_{2}$ represents solenoidal acceleration because 
$({\VEC{\nabla}}\times \rho_{0} ^{-1} {\VEC{f}})_{\phi}$
$=Q_2 (\rho_0 ^{-1} )^\prime r^{-1} P_{2,\theta}$.
Equivalently, the term $Q_{2}$ represents the non-barotropic structure.
The following equation, which 
is proportional to $Q_{2}$ is zero for the barotropic case;
\begin{equation}
{\VEC{\nabla}}(\rho_{0} +\delta \rho_{2}P_{2})
\times {\VEC{\nabla}}(p_{0} +\delta p_{2}P_{2})
\approx \frac{\rho_{0} ^\prime Q_2}{r} P_{2,\theta} {\VEC{e}}_{\phi}.
\end{equation}
The relation (\ref{p2withq2.eqn}) is reduced to 
$\delta p_{a2} = c_s ^2 \delta \rho_{a2}$ for $Q_2=0$.
The function $Q_2$ is proportional to the normalization constant $N_{n}$ because
$A_{2} \propto N_{n}$ and $B_{2} \propto N_{n}$.
We introduce a non-dimensional scaling parameter $\zeta$,
such that the maximum of $|Q_2 /p_0|$ is $\zeta$.
A reasonable range of $\zeta$ is $0 < \zeta <1$ because
$|\delta p_{a2}/p_0| \sim  |Q_2 /p_0| \sim \zeta$.
Hereafter, we use $\zeta$ instead of $N_{n}$.
The Poisson equation (\ref{PoissonA.eqn}) for $l=2$ is 
\begin{equation}
(r^2 \delta \Phi_{a2} ^\prime)^\prime-6
    \delta \Phi_{a2}= 4\pi G_{\rm{N}} r^2 \delta \rho_{a2}.
    \label{poissonrd_l2.eqn}
\end{equation}
We assume a spherical core ($r< r_c$) and a non-spherical perturbation is localized at its outer side ($r_c \le r \le R$). 
The boundary conditions for $\delta \Phi_{2}$ are as follows:
\begin{equation}
  \delta \Phi_{a2}(r_c) ^\prime =\frac{2}{r_c} \delta\Phi_{a2}(r_c),  
    ~~
  \delta \Phi_{a2}(R) ^\prime =-\frac{3}{R} \delta \Phi_{a2}(R).
  \label{poissonBC1.eqn}
\end{equation}
The solution of eq.~(\ref{poissonrd_l2.eqn}) at $R$ is given by
$\delta \Phi_{a2}  =- G_{\rm{N}}{\mathcal{Q}}_{a} /R^3$,
where ${\mathcal{Q}}_{a}$ characterizes the quadrupole moment obtained by the following integration:
\begin{equation}
     {\mathcal{Q}}_{a} \equiv \frac{4\pi}{5}
     \int_{r_c} ^{R} \delta \rho_{a2}  r^4 dr.
\end{equation}
The relation for the quadrupole moment $Q_{22}$ used in gravitational-wave studies is
${\mathcal{Q}} =({8\pi}/{15})^{1/2}  Q_{22}$, where
the coefficient $({8\pi}/{15})^{1/2}$ is obtained from the normalization of the angular function.
In the latter, the density perturbation is expanded as
$\delta \rho = {\rm{Re}}$
$[\sum _{lm} \delta {\hat \rho}_{lm} Y_{lm}(\theta,\phi)]$.
The spin axis is orthogonal to the symmetric axis of our model.
The value $ {\mathcal{Q}}_{a}$ is numerically obtained 
for Models I--III as 
\begin{align*}
& {\mathcal{Q}}_{a} /\zeta = 3.397 \times 10^{-4} M R^2  =  
 1.468 \times 10^{42}~~
 {\text{g~cm}}^2~~
 ({\text{Model I}}),
 \\
& {\mathcal{Q}}_{a}  /\zeta 
= 7.168 \times 10^{-5} M R^2 = 
3.097 \times 10^{41} ~~ 
  {\text{g~cm}}^2~~
({\text{Model II}}),  
\\
&  {\mathcal{Q}}_{a} /\zeta= 9.589 \times 10^{-6} M R^2 = 
 4.142 \times 10^{40} ~~ 
  {\text{g~cm}}^2~~
({\text{Model III}}),  
\end{align*}
where $\zeta$ is a scaling factor that characterizes the magnitude of the perturbation. 
The magnitude of ${\mathcal{Q}}_{a}$ varies by more than one order, depending on the distribution of the perturbation.
In addition, we changed the ratio of $A_2$ to $B_2$ 
in eq.~(\ref{explictradfnH.eqn}), and
calculated ${\mathcal{Q}}_{a}$ for Model I.
The results are 
${\mathcal{Q}}_{a} /(\zeta M R^2) = 6.844 \times 10^{-4}$ for $A_2 =10B_2$, and
${\mathcal{Q}}_{a} /(\zeta M R^2) = 2.996 \times 10^{-4}$ for $B_2 =10A_2$.
These values are not different from that of $A_2 =B_2$;
hence, the ratio is not very important.
Using the solution $\delta \Phi_{a2}$ of eq.~(\ref{poissonrd_l2.eqn}),
the radial functions $\delta \rho_{a2}$ (\ref{fld_dns_phi.eqn})
and $\delta p_{a2}$ (\ref{fld_pr_phi.eqn}) are shown in Fig.~\ref{Fig2}.
They are obtained by multiplying a factor $\zeta^{-1}$ because the perturbation amplitudes are proportional to $\zeta$.
The radial profile of $\delta \Phi_{a2}$ is approximately a sine curve
for all models; however, the spatial range differs depending on the assumed force.
The non-zero range causes a significant difference in the integrated value.
The value ${\mathcal{Q}}_{a}$ in Model I is the largest because the
perturbation of the mass density is distributed over the crust.
In the right panel of Fig.~\ref{Fig2},
 $\delta p_{a2}$, $c_{s} ^{2} \delta \rho_{a2}$, and $Q_{2}$ are shown.
The resultant $\delta p_{a2}$ ($=c_{s} ^{2} \delta \rho_{a2} +Q_{2}$)
differs significantly from $c_{s} ^{2} \delta \rho_{a2}$.
Our deformed models cannot be described by the adiabatic change 
$c_{s} ^{2} \delta \rho_{a2}$ from the background model. 
The additional term $Q_{2}$ is an important factor in deformation.

The maximum value of the function $|Q_{2}(r)|/p_{0}(r)$
is set as $\zeta$ in our calculation.
The maximum of $|Q_{2}|/p_{0}$ is not the ''maximum'' of $Q_{2}/p_{0}$ at $r_1$, but
the ''minimum'' at $r_2$ ($r_1<r_2$), i.e., 
the maximum of $|Q_{2}|/p_{0}$ is $-Q_{2}(r_2)/p_{0}(r_2)$.
Figure~\ref{Fig2} may be confusing because $|Q_{2}(r_1)| > |Q_{2}(r_2)|$.
Note that $p_{0}$ is a steep decreasing function.
This normalization affects the overall magnitude of $Q_{2}$,
which decreases with $r_2$, as shown in Fig.~\ref{Fig2}.

  \section{Residual deformation sustained by the elastic force}
We study the equilibrium state B after solidification, in which the driving force has already vanished.
We solve eq.~(\ref{balanceAF.eqn})--(\ref{PossionPT.eqn}) by expanding the displacement vector ${\VEC{\xi}}$
and other scalar functions in terms of the Legendre polynomial of $l=2$; 
\begin{equation}
\VEC{\xi} =  r R_{2}(r) P_{2} \VEC{e}_{r} +
r U_{2}(r)  P_{2,\theta} \VEC{e}_{\theta}.
\end{equation}
The perturbation equations are derived by neglecting the gravitational perturbation in previous studies~\citep[see e.g.,][]
{1988ApJ...325..725M,2000MNRAS.319..902U,2024ApJ...974..125K}.
They are described by a set of fourth-order differential equations; however, they
are written using different functions and variables.
Our formalism extends a set of equations in \cite{2024ApJ...974..125K} by accounting for additional terms
$\VEC{f}$ and $Q$ as well as gravitational perturbation.
By eliminating $\delta \rho_{b}$ and $\delta p_{b}$ with eqs.~(\ref{densitybyxi.eqn}) and (\ref{presbwithQ.eqn}), 
the perturbation equations~(\ref{balanceAF.eqn})--(\ref{PossionPT.eqn}) are reduced to the following set of sixth-order differential equations:
\begin{align}
 &R_{2} ^{\prime} = \frac{6}{r} U_{2}
 -\left( 3+ 
 \frac{r p_{0} ^{\prime}}{\Gamma p_{0}} \right)\frac{R_{2}}{r}+
 \frac{1}{\Gamma p_{0}r^4} \left(
 X_{2} -r^3 Q_{2} \right),
 \label{plRUXY1.dfeqn}
  \\
&U_{2} ^{\prime} = -\frac{R_{2}}{r} + \frac{Y_{2}}{\mu r^4},
 \label{plRUXY2.dfeqn}
  \\
&X_{2} ^{\prime} = \frac{6}{r} Y_{2}
+\left( 3+ 
 \frac{r p_{0} ^{\prime}}{\Gamma p_{0}} \right) 
 \left(
 \frac{X_{2}}{r} -12 \mu r^2 U_{2}  \right)
    \nonumber\\
  &~~~~  - \left[
  (\rho_{0} ^{\prime} g +4\pi G_{\rm{N}} \rho_{0} ^2)r^2 +
   \frac{(rp_{0} ^{\prime})^2}{\Gamma p_{0}}
  - 4 \mu \left( 3 + 
 \frac{2 r p_{0} ^{\prime}}{\Gamma p_{0}} 
  \right) \right]r^2 R_{2} 
    \nonumber\\
  &~~~~ + \rho_0 r W_{2}
 +\left( \frac{4\mu}{\Gamma p_{0} }-\frac{r p_{0} ^{\prime}}{\Gamma p_{0}}
 \right)r^2Q_{2} +\rho_{0} r^3 A_{2},
     \label{plRUXY.dfe3qn}
    \\
&Y_{2} ^{\prime} = -
\frac{ X_{2}}{r} + 22\mu r^2 U_{2} 
-\left(6 + 
 \frac{2r p_{0}^{\prime}}{\Gamma p_{0}} \right)\mu r^2R_{2} 
    \nonumber\\
    &~~~~ +\rho_0 r^2 \delta \Phi_{b2}
  -\frac{2\mu r^2}{\Gamma p_{0}}Q_{2} +\rho_{0} r^3 B_{2}.
      \label{plRUXY4.dfeqn}
  \\
&W_{2} ^\prime = 6(  \delta \Phi_{b2}+
4 \pi G_{\rm{N}} r^2 \rho_{0}U_{2}),
\label{poisson1.firstODE}
\\
&\delta \Phi_{b2} ^\prime =
\frac{W_{2} }{r^2}- 4 \pi G_{\rm{N}} r\rho_{0} R_{2},
\label{poisson2.firstODE}
 \end{align}
where the adiabatic constant is $\Gamma =1.5$.
We use the condition $\mu \ll p_{0}$ to simplify 
eqs.~(\ref{plRUXY1.dfeqn})--(\ref{plRUXY4.dfeqn})
based on the equations in \cite{2024ApJ...974..125K}.
The function $W_2$ is introduced into the Poisson equation for reduction to first-order differential equations
(\ref{poisson1.firstODE})--(\ref{poisson2.firstODE}).
The functions $X_2$ and $Y_2$ represent traction for the radial and perpendicular components, respectively. 
The traction must be continuous across the surfaces at $r= r_c$ and $R$.
The shear stress vanishes outside the crust.
Conditions 
$\sigma_{rr} = 2(r R_{2}^\prime + 3U_{2} )P_{2}/3 =0$ and 
$\sigma_{r \theta} =(r U_{2}^\prime + R_{2}) P_{2,\theta}/2 =0$
are imposed to ensure continuity.
By using eqs.~(\ref{plRUXY1.dfeqn})--(\ref{plRUXY2.dfeqn}),
the conditions at $r= r_c$ and $R$ lead to the following algebraic ones:
\begin{equation}
    X_{2} -(rp_{0} ^{\prime} +3\Gamma p_{0})r^3R_{2} 
  +9\Gamma p_{0} r^3 U_{2}=Q_{2}r^3,
  ~~~~
 Y_{2}=0. 
\end{equation}
The boundary conditions for the Poisson equation are
the same as those in eq.~(\ref{poissonBC1.eqn}), which are written as 
\begin{align}
  & W_{2}- 2r_c \delta\Phi_{b2}
 -4\pi G_{\rm{N}} \rho_{0}r_c ^3 R_2=0 ~~~{\text{at}~r_c} ,
 \nonumber
 \\
 &W_{2} + 3R \delta \Phi_{b2}
 -4\pi G_{\rm{N}} \rho_{0} R^3 R_2=0 ~~~{\text{at}~R}. 
\end{align}
The solution at $R$ is given by
$\delta \Phi_{b2}  =- G_{\rm{N}} {\mathcal{Q}}_{b} /R^3$,
where 
\begin{equation}
     {\mathcal{Q}}_{b} \equiv \frac{4\pi}{5}\int_{r_c} ^{R} \delta \rho_{b2}  r^4 dr.
\end{equation}
The numerical calculations yield 
\begin{align*}
& {\mathcal{Q}}_{b} /\zeta = -3.332 \times 10^{-4} M R^2  =  
  -1.439 \times 10^{42}~~
 {\text{g~cm}}^2~~
 {\text{(Model I)}},
 \\
& {\mathcal{Q}}_{b}  /\zeta 
= -7.069 \times 10^{-5} M R^2 = 
-3.054 \times 10^{41} ~~ 
  {\text{g~cm}}^2~~
{\text{(Model II)}},  
\\
&  {\mathcal{Q}}_{b} /\zeta= -9.221 \times 10^{-6} M R^2 = 
 -3.983 \times 10^{40} ~~ 
  {\text{g~cm}}^2~~
{\text{(Model III)}}.  
\end{align*}
The results show that
${\mathcal{Q}}_{b}\approx - {\mathcal{Q}}_{a}$; however, ${\mathcal{Q}}_{a}+{\mathcal{Q}}_{b}$
is positive in each model.
The residual ${\mathcal{Q}}_{a}+ {\mathcal{Q}}_{b}$ is 1.4--3.8\% of the original
${\mathcal{Q}}_{a}$.
%

\begin{figure}\begin{center}
 \includegraphics[width=1.0\columnwidth]{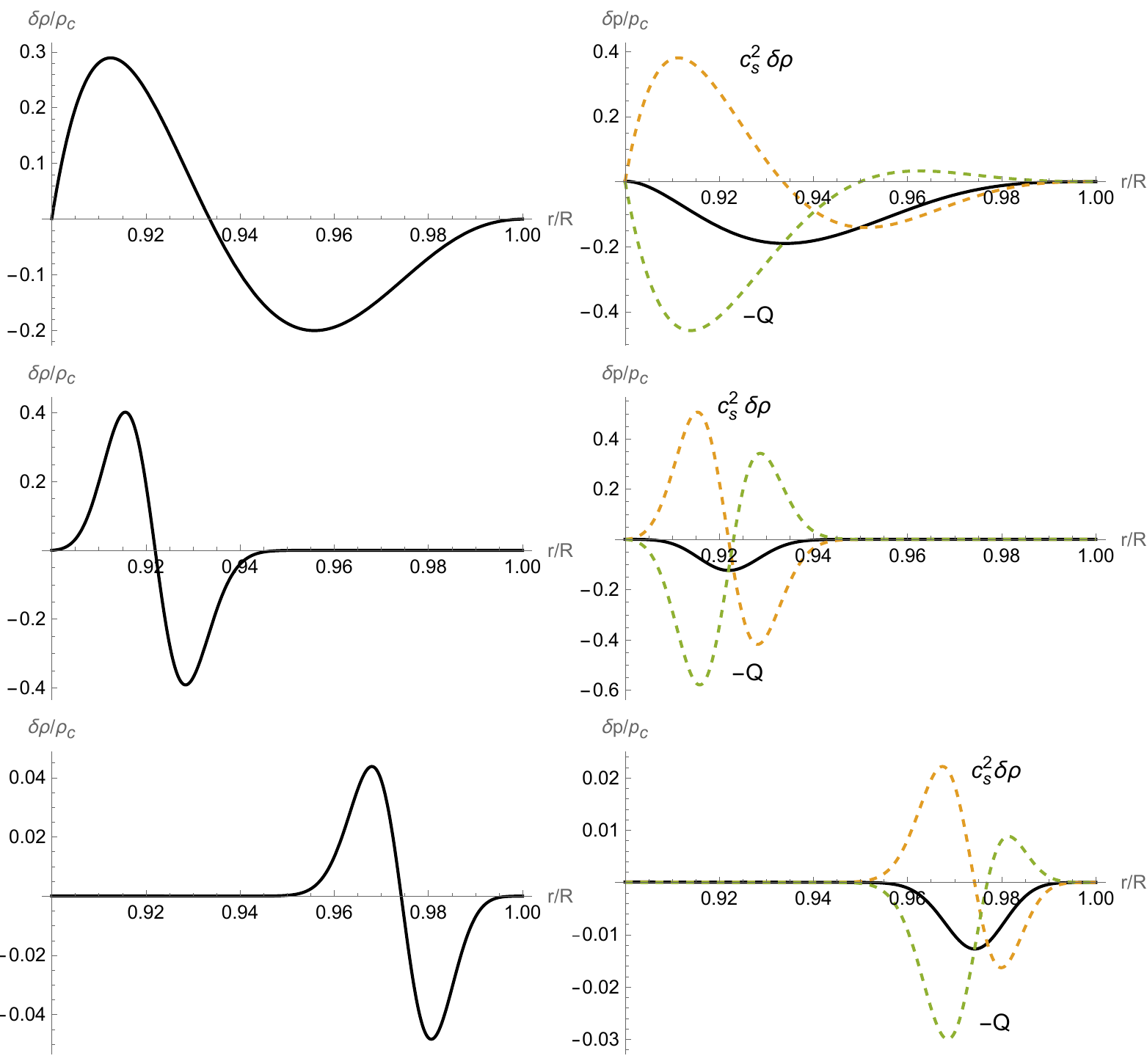}
\caption{ 
\label{Fig3}
$\zeta^{-1}\delta \rho_{b2}$ (left panel) and $\zeta^{-1}\delta p_{b2}$ (right panel)
as a function of the radius.
In the right panels, 
$-\zeta^{-1}Q_{2}$ and $\zeta^{-1} c_s ^2 \delta \rho_{b2}$ are also shown by dashed curves.
From top to bottom, the models correspond to I, II, and III.
}
\end{center}\end{figure}

%
The radial functions $\delta \rho_{b2}$ and $\delta p_{b2}$ are
numerically calculated by the solutions of 
eqs.~(\ref{plRUXY1.dfeqn})--(\ref{poisson2.firstODE}).
Figure~\ref{Fig3} shows the results.
The relationship between $\delta \rho_{a2}$ and $\delta \rho_{b2}$ is not trivial.
However, a comparison of Figs.~\ref{Fig3} and ~\ref{Fig2} shows a remarkable fact: 
$\delta \rho_{b2}$ is approximately the same as $\delta \rho_{a2}$ in magnitude, but opposite in sign.
The relationship between $\delta p_{b2}$ and $\delta p_{a2}$ is also the same.
The deviations $\delta \rho_{a2}$ and 
$\delta p_{a2}$ from a spherical background are driven by $+\VEC{f}$, whereas
$\delta \rho_{b2}$ and $\delta p_{b2}$ are driven by $-\VEC{f}$.
Therefore, the opposite sign is reasonable.
The coincidences in the profile and magnitude, such as
$\delta \rho_{b2} \approx -\delta \rho_{a2}$, are interesting
because they are produced in different systems,
in which the elastic force acts or does not.
The sum $\delta \rho_{a2}+\delta \rho_{b2}$ is not exactly zero; however, it is a small value. 
We further discuss it in terms of the integral
${\mathcal{Q}}$ in the next section.

\begin{figure}\begin{center}
\includegraphics[width=1.0\columnwidth]{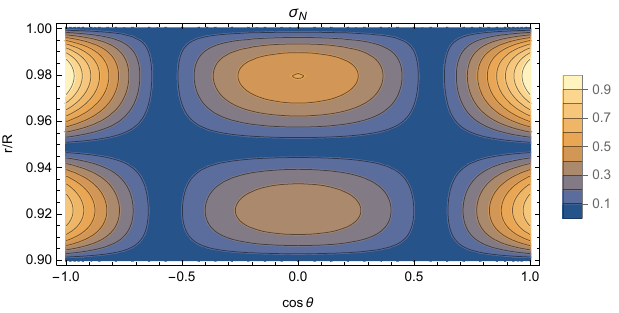}
\includegraphics[width=1.0\columnwidth]{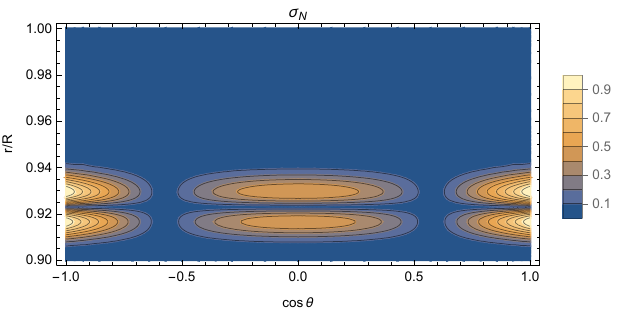}
\includegraphics[width=1.0\columnwidth]{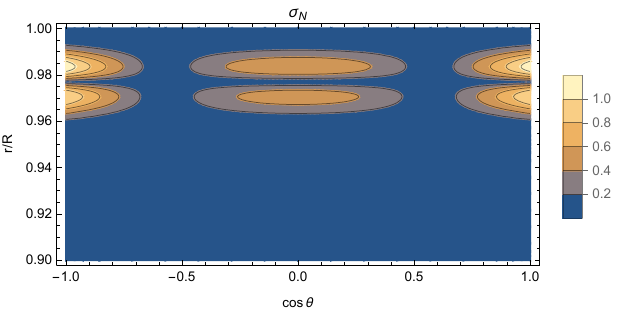}
\caption{ 
\label{Fig4}
The magnitude of the normalized shear strain $\sigma/\sigma_{\rm{max}}$ is shown 
by contours in the meridian plane.
The models correspond to I, II, and III from top to bottom.
}
\end{center}\end{figure}


%
The shear stress $\sigma_{ij}$ is calculated from ${\VEC{\xi}}$.
Figure~\ref{Fig4} shows the contours of the magnitude
$\sigma \equiv (\sigma_{ij}  \sigma^{ij}/2 )^{1/2}$. 
The peak of $\sigma$ clearly traces the magnitude of the driving force.
This contrasts with the sharp peak near the surface in previous studies~\citep{2021MNRAS.500.5570G,2022MNRAS.517.5610M}.
The angular dependence of $\sigma$ is due to $P_{2}(\cos \theta)$, because
$\sigma_{rr}$, which is related to $-(\sigma_{\theta \theta}+\sigma_{\phi \phi})$ by
the traceless property, dominates.
In equilibrium at stage B, the diagonal part of the shear stress
$\sigma_{ij}$ is replaced by the decreasing pressure term.
The maximum in the crustal region for three models is obtained as 
\begin{equation*}
 \sigma _{\rm{max}} =0.41\zeta~~  
 ({\text{Model I}}),
~~~
   0.39\zeta ~~ 
({\text{Model II}}),  
~~~
   0.34\zeta ~~ 
({\text{Model III}}).  
\end{equation*}
The elastic equilibrium of the crust breaks down when 
$\sigma > \sigma _{c}$(the von Mises criterion),
where the critical number is $\sigma_c \approx 10^{-2} -10^{-1}$
\citep{2009PhRvL.102s1102H,2018MNRAS.480.5511B,2018PhRvL.121m2701C}.
The condition is unlikely to be satisfied for $\zeta < 0.1$.
Thus, a deformed crust may still exist.

  \section{Ellipticity}
A quadrupole deformation is characterized by a dimensionless quantity;
\begin{equation}
     \epsilon  = \frac{I_{xx}-I_{yy}}{I_{zz}}
     =\frac{5}{2}\frac{{\mathcal{Q}}}{MR^2} .
%
\end{equation}
The quadrupole moment is approximately determined by the spherical core, and the denominator is approximated using $2MR^2/5$ 
$=1.73\times 10^{45} {\text{g~cm}}^2$ as the uniform density of a star with mass $M$ and radius $R$.
However, the difference in the numerator is given by the deformed crust.
The deformations before crustal solidification for the three models are obtained using
${\mathcal{Q}}_{a}$;
\begin{equation*}
 \epsilon_a = 8.493 \times 10^{-4}\zeta~
({\text{I}}),
~~
   1.792 \times 10^{-4}\zeta ~
 ({\text{II}}),
\\
~~
   2.397 \times 10^{-5}\zeta ~
({\text{III}}).
%
\end{equation*}
The relic deformations using
${\mathcal{Q}}_{a}+{\mathcal{Q}}_{b}$ are
\begin{equation*}
 \epsilon_a +\epsilon_b = 
 1.6 \times 10^{-5}\zeta~ 
 ({\text{I}}),
~~
2.4 \times 10^{-5}\zeta ~
 ({\text{II}}),
~~
9.2 \times 10^{-7}\zeta ~ 
({\text{III}}).
%
\end{equation*}
%

\begin{figure}\begin{center}
  \includegraphics[width=0.9\columnwidth]{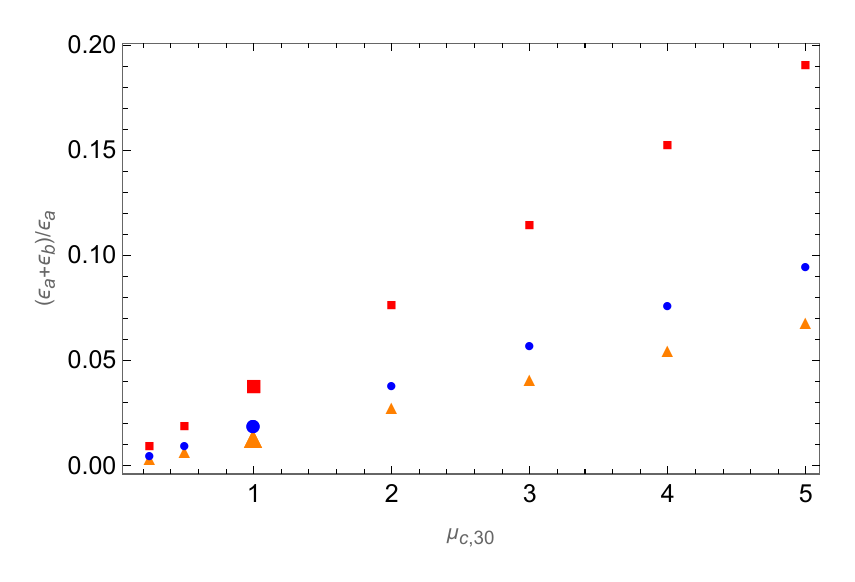}
\caption{ 
\label{Fig5}
Fraction $(\epsilon _{a}+\epsilon _{b})/\epsilon _{a}$
 as a function of the shear modulus $\mu_{c,30}\equiv \mu_c/(10^{30} {\text{erg~cm}}^{-3})$.
Dots, triangles, and squares represent Models I, II, and III, respectively.
}
\end{center}
\end{figure}


We consider how the overall normalization factor $\mu_c$ in eq.~(\ref{DFshMD.eqn})
changes the residual.
Figure~\ref{Fig5} shows the fraction of the residual $(\epsilon_{a}+\epsilon_{b})/\epsilon_{a}$.
For the fiducial value $\mu_c=10^{30} {\text{erg~cm}}^{-3}$, 
the sum $\epsilon _{a}+\epsilon _{b}$ is a small percentage of $\epsilon _{a}$.
However, the residual increases with $\mu_c$ for all models.
This implies that the reduced deformation in the latter state B occurs owing to a weaker elastic force.
By extrapolating the results, the deformation would be sustained at the same magnitude if
$\mu_c \sim 10^{32} {\text{erg~cm}}^{-3}$, which is unrealistic.
We discuss an allowed range of $\zeta$, which represents the maximum value of $Q_{2}/p_0$, and
the magnitude of the variations in density $\delta p_{a2}/p_0$ and pressure 
$\delta \rho_{a2}/\rho_0$.
The solid crust forms when the temperature decreases below the threshold;
it ranges from $T\sim 6 \times 10^{9}$K at the core--crust interface 
to $T\sim 6 \times 10^{8}$K in the outer region.
This value is below the Fermi temperature $T_{\rm{F}}(\approx10^{11-12}~{\rm K})$ for both
relativistic electrons and non-relativistic neutrons.
Taking the Sommerfeld approximation, thermal pressure $p_{\rm{th}}$ is of the order
$p_{\rm{T}}/p_{0} \sim \alpha_{n}(T/T_{\rm{F}})^2$,
where $\alpha_{\rm{Rel}} =2\pi^2$ for electrons and $\alpha_{\rm{NR}} =5\pi^2/8$
for neutrons.
At the onset of solidification near the core at $T\simeq T_{c} \approx 6.4\times 10^{9}$K,
$p_{\rm{T}}/p_{0} \sim 2\times 10^{-3}$ owing to thermal pressure by neutrons near the core, and 
$p_{\rm{T}}/p_{0} \sim 2\times 10^{-2}$ owing to that by electrons near the surface\footnote{Recently, \cite{2025MNRAS.540.2349J} suggested the importance of the finite temperature effect of the lattice pressure in the crust;
the pressure is comparable to the neutron gas pressure around $T\approx$ a few times $10^8~{\rm K}$
because the former behaves as $\propto T^4$ in a low-$T$ regime. However, the temperature in our case is 1--2 orders of magnitude higher than they considered. The lattice pressure is approximated by classical thermodynamics, i.e., $\propto T^1$. As such, it is smaller than the degenerate gas pressure.}
Without non-spherical fluctuation of the temperature distribution, these estimates of $p_{\rm{T}}/p_{0}$ introduce a constraint of $\zeta<10^{-2}$, providing an optimistic estimate of the relic ellipticity $\epsilon_a +\epsilon_b\sim10^{-7}$.
Notably, two uncertain factors remain for obtaining a more precise constraint on $\zeta$. One is the value $T_c$, since it depends on the equation of state at subnuclear densities. According to Fig. 1 in \cite{2025PTEP.2025l3E01N}, the critical temperature of solidification increases by up to an order of magnitude, leading to an increase in $p_{\rm{T}}/p_{0}$ or the upper limit value of $\zeta$ by up to two orders of magnitude. The other is the spatial fluctuation of the temperature distribution, although quadrupole fluctuation is often \textit{assumed}.
Future studies investigating the structure and thermal evolution of the hot crust are necessary to verify our CGW observation scenario.

  \section{Summary and discussion}
In this study, we investigate the fate of quadrupole deformation produced in a hot era.
The magnitude generally decreases as the driving force decays; however, the residual sustained by the elastic force is finite.
The deformation is imprinted on the solid crust through the solidification era.
We demonstrate that the relic ellipticity is a small percentage of the initial one.
The significant reduction rate originates from the weakness of the
elastic force compared with the hydrodynamical force that sustained the deformation prior to solidification.
A similar ellipticity reduction was also calculated in a previous study ~\citep{2021MNRAS.500.5570G,2022MNRAS.517.5610M}.
However, the residual ellipticity ratio ranged from $10^{-6}$ to $10^{-1}$, depending on the assumed force.
These values are different from our results.
The deformation of the entire star has been calculated in previous studies, with a drastic change in the stellar core observed in some cases.
Our calculation is limited to the crustal part only; hence, 
the reduction rate does not vary significantly.
The solidification effect is easily understood in our simple models.
Our model also shows an interesting spatial location of relic deformation, which is clearly traced by shear strain.
This profile also contrasts with that in the previous work, in which
a sharp peak is located near the surface, irrelevant to the driving force.
The relic remains within a stable range because the von Mises criterion is satisfied, 
unless the initial deformation is extremely large.
Thus, the deformation in the elastic crust remains without fracture.
In this study, we explore the evolutionary reduction of deformation. However, the relic deformation relevant to gravitational-wave astronomy remains unclear owing to its dependence on the initial value. The crucial unknown factor is the ellipticity around $10^2$--$10^4$ s at the hot era of a neutron star.
The magnitude in the past is one or two orders of magnitude larger than the observable relic. 
Furthermore, the deformation mechanism and magnitude in the early history of the neutron star
are unknown.
Our model of the deformation and background crust may be rather simple, and therefore, the magnitude of ellipticity
produced in the hot era is relatively less reliable.
Smooth deformation in a large scale considered in this paper may be altered by a detailed model of the local structure.
Our concern of this study by the simple models is to demonstrate the relic deformation that remains in the solid crust. 
Successful detection in CGWs will prove the hot state of a neutron star
before solidification.
This study provides the first step toward exploring the link between 
potential deformation in past and future observations.
However, further studies are necessary to reach a definite conclusion.

\section*{Acknowledgments}
 This work was supported by JSPS KAKENHI Grant Numbers JP23K03389 (YK), JP25K17403 (AD), JP22K03681 and JP23K22538 (SK).

 \section*{DATA AVAILABILITY}
The data underlying this article will be shared on reasonable request to the corresponding author.

\bibliographystyle{mnras}
\bibliography{kojima26Jan}

\bsp	
\label{lastpage}
\end{document}